\documentclass[sigconf,nonacm]{acmart}
\AtBeginDocument{%
  }

\usepackage{threeparttable}
\usepackage{multirow}
\usepackage{multicol}

\begin{document}


\title{Automated Feedback on Student-Generated UML and ER Diagrams Using Large Language Models}

\author{Sebastian Gürtl}
\email{sebastian.guertl@tugraz.at}
\orcid{0009-0006-9008-7147}
\affiliation{%
  \institution{Graz University of Technology}
  \city{Graz}
  \country{Austria}
}

\author{Gloria Schimetta}
\email{gloria.schimetta@student.tugraz.at}
\orcid{0009-0003-7824-0805}
\affiliation{%
  \institution{Graz University of Technology}
  \city{Graz}
  \country{Austria}
}

\author{David Kerschbaumer}
\email{david.kerschbaumer@tugraz.at}
\orcid{0009-0007-1861-9417}
\affiliation{%
  \institution{Graz University of Technology}
  \city{Graz}
  \country{Austria}
}

\author{Michael Liut}
\orcid{0000-0003-2965-5302}
\email{michael.liut@utoronto.ca}
\affiliation{%
  \institution{University of Toronto Mississauga}
  \city{Mississauga}
  \country{Canada}
}

\author{Alexander Steinmaurer}
\email{alexander.steinmaurer@it-u.at}
\orcid{0000-0002-1760-2855}
\affiliation{%
  \institution{Interdisciplinary Transformation University Austria}
  \city{Linz}
   \country{Austria}
}


\renewcommand{\shortauthors}{Sebastian Gürtl, Gloria Schimetta, David Kerschbaumer, Michael Liut, and Alexander Steinmaurer}
\begin{abstract}
UML and ER diagrams are foundational in computer science education but come with challenges for learners due to the need for abstract thinking, contextual understanding, and mastery of both syntax and semantics. These complexities are difficult to address through traditional teaching methods, which often struggle to provide scalable, personalized feedback, especially in large classes. We introduce DUET (Diagrammatic UML \& ER Tutor), a prototype of an LLM-based tool, which converts a reference diagram and a student-submitted diagram into a textual representation and provides structured feedback based on the differences. It uses a multi-stage LLM pipeline to compare diagrams and generate reflective feedback. Furthermore, the tool enables analytical insights for educators, aiming to foster self-directed learning and inform instructional strategies. We evaluated DUET through semi-structured interviews with six participants, including two educators and four teaching assistants. They identified strengths such as accessibility, scalability, and learning support alongside limitations, including reliability and potential misuse. Participants also suggested potential improvements, such as bulk upload functionality and interactive clarification features. DUET presents a promising direction for integrating LLMs into modeling education and offers a foundation for future classroom integration and empirical evaluation.

\end{abstract}




\maketitle

\section{Introduction}
Software modeling is an integral part of software development, system design, or business process modeling. It requires skills in system analysis, abstraction, and problem solving, but also knowledge of modeling languages. Two of these widely used notations are the \textit{Unified Modeling Language} (UML) and \textit{Entity Relationship} (ER) diagrams.

Although UML is treated as standard for software engineering, its application in the computer science (CS) industry has dwindled and only selected components, such as class and sequence diagrams being used for high-level conceptualization, have remained \cite{6606618, romeo2025uml}.

Despite this lack of use in the industry, UML is believed to be beneficial in CS education, as its concepts and notation provide a helpful framework for novice students \cite{6606618}. However, learners often struggle with the complexity of UML and frequently make mistakes in notation, scope, and diagram interpretation \cite{akayama2013tool,8802110,erickson2007can,9125110}. Ideally, these challenges require guidance and feedback from educators and teaching assistants (TAs) while students learn.  

Traditionally, undergraduate CS classes are characterized by a large number of students, which stands in contrast to the need for individual feedback. For this reason, feedback and assessment often rely on rigid, rule-based methods that lack personalized guidance. Large Language Models (LLMs), on the other hand, excel in natural language understanding, pattern recognition, and adaptive feedback generation, making them well-suited for analyzing student diagrams and offering targeted suggestions \cite{Rivera:2024}. In addition, LLMs can engage in interactive dialogues to clarify design goals and constraints, ultimately improving both productivity and design quality \cite{ahmad2023towards,de2023echo}. The growing popularity of prompt engineering further enhances the potential of LLMs, leading to more effective and tailored results in software design feedback \cite{sengul2024software}.

To address this, we propose an LLM-based tool that transforms UML and ER diagrams into PlantUML -- a tool that textually represents diagrams. The tool subsequently analyzes these diagrams by comparing them against a reference solution provided by the instructor. Based on this, it generates hints and improvement suggestions for the students while the instructors can view common issues and teaching recommendations.

\section{Background and Related Work}
\subsection{UML and ER Diagrams in CS Education}
While practitioners in the industry have mostly abandoned UML, it is still highly relevant in introductory CS and software engineering classes \cite{6606618}. The process of visualizing system architecture and data structures is challenging for novice software engineers. UML and ER diagrams enforce the principles of software abstraction and guide developers in structuring complex systems. Students, however, struggle with the complexity of UML and ER notation, leading to frequent mistakes \cite{akayama2013tool}. These mistakes have been extensively discussed in literature \cite{8802110, 9125110}, with the following being prominent in UML diagrams: 
\begin{itemize}
    \item General Difficulties: Diagram construction and symbol use.
    \item Class Diagram: Attributes, inheritance confusion.
    \item Sequence Diagram: Lifeline naming, object interaction.
\end{itemize}

The most common issues of ER diagrams include:
\begin{itemize}
    \item Inconsistency with the class diagram.
    \item Missing or redundant relationships.
    \item Wrong multiplicities.
\end{itemize}

Due to the variety of tools available, students can take different approaches to creating a UML or ER diagram, including textual representations.

\subsection{PlantUML}
PlantUML\footnote{\url{https://plantuml.com/en/}} is a popular tool that allows the creation of various types of diagrams, including most UML diagrams, as well as ER diagrams. Romeo et al. \cite{romeo2025uml} analyzed 13,000 open-source repositories on GitHub and found that PlantUML has been the most used text-based UML tool since UML's resurgence in 2016. Since LLMs still lack the ability to extract high-detail information from images effectively, text representations of UML diagrams, such as PlantUML, can be more accurately processed and analyzed.

\subsection{LLMs for UML and ER Diagram Generation}

LLMs are trained on vast datasets for various tasks. Given the advantages of text-based UML representations, recent research has explored how LLMs can assist in generating, interpreting, and optimizing PlantUML diagrams. While not explicitly trained for diagram generation, they can still assist students in UML modeling by providing examples and naming suggestions. However, LLMs have inherent limitations as shown in various studies \cite{wang2024llms,conrardy2024image,camara2023assessment,wang2024assessing}.

Wang et al. \cite{wang2024llms} conducted a study analyzing UML models created by students who used OpenAI's ChatGPT as a support tool for their modeling tasks. The study found that it correctly identified 66\% of classes, 75\% of operations, and 91\% of attributes. However, it struggled with class relationships, achieving only 25\% accuracy. 
LLMs, particularly GPT-4, perform notably better with sequence diagrams, correctly identifying 74\% of objects, 68\% of messages, and maintaining the correct message order 82\% of the time. 

Multimodal LLMs still face challenges in analyzing diagrams, particularly UML and ER diagrams. In case studies by Conrardy et al. and Camara et al. \cite{conrardy2024image, camara2023assessment}, GPT-4 significantly outperformed both open-source and commercial multimodal LLMs in converting hand-drawn UML diagrams to PlantUML. In one experiment, GPT-4 did not produce a single syntax error after 36 generations, while other models produced multiple. Nonetheless, all models, including GPT-4, made minor mistakes requiring human correction. GPT-4 outperformed all the other LLMs, often producing near-perfect results in a single iteration.

Another study published by Wang et al. \cite{wang2024assessing} compared GPT-4's UML grading to that of a human instructor. Its assessments matched the instructor's in approximately 50\% of cases. Discrepancies were attributed to  \textit{misunderstanding} (e.g., the grading schema), \textit{overstrictness} (e.g., rejecting similar valid alternatives), and \textit{wrong identification} (e.g., GPT-4 failed to recognize a relationship). The study also confirmed that sequence diagrams are less prone to errors in GPT-4's assessments. Across 40 UML diagrams (40 each), \texttt{GPT-4o} correctly evaluated approximately 35 class and sequence diagrams, making three incorrect identifications for class diagrams but none for sequence diagrams.

As seen in the studies above, LLMs show promise in analyzing and generating UML and ER diagrams. However, automated evaluation tools (even without the use of LLMs) can provide additional structured feedback that can further support students in their learning process.

\subsection{Automated Evaluation of UML and ER Diagrams}
To address the challenges students have experienced while working with modeling languages, several evaluation tools have been developed to assist with the learning process.

\textit{UML Miner} \cite{10350741} applies process mining to study student behavior in Visual Paradigm\footnote{\url{https://www.visual-paradigm.com}}, performing conformance checking against expected behavior and similar solutions. This approach, however, relies on manual feedback from educators, and the students do not receive direct feedback from the tool.

\textit{UML Mentor} \cite{10.1145/3649409.3691078} introduces a peer-review system where students create UML diagrams for software design challenges and receive feedback from peers. While it eases instructor workload, peer-feedback may lack expertise and is limited to predefined tasks.

\textit{UMLegend} \cite{garaccione2025gamification} incorporates gamification, creating a sandbox environment for students to solve provided exercises. Live feedback notifies the student of errors. This technique proved to increase diagram correctness 11\% in an experiment. UMLegend, however, limits students in the creation of the diagram -- hand-drawn diagrams as well as diagrams drawn in other tools cannot be evaluated. The evaluation engine is restricted in its assessment capabilities, as it relies on a single reference solution for comparison.

In summary, despite UML's declining industry use, it remains an essential educational tool. Students often struggle with UML and ER diagrams, necessitating automated evaluation solutions. While LLMs, particularly GPT-4, demonstrate promising results in analyzing and generating UML and ER diagrams, existing evaluation tools demonstrate that human oversight remains essential in the near future to guide LLM-based assessments. Meanwhile, tools that do not leverage LLMs provide alternative feedback approaches, offering valuable insights for alternative implementations.

\section{Diagrammatic UML \& ER Tutor}

Building on the motivation, challenges, and related work, this section introduces the proposed automated feedback tool \textit{Diagrammatic UML \& ER Tutor} (DUET) designed to address educational gaps. In particular, we present the tool's design, system architecture, workflow, and user interaction. It utilizes advanced LLMs, such as GPT-4o, and smaller complementary LLMs to provide meaningful automated feedback on student-generated UML and ER diagrams. Students upload self-drawn diagrams, which are compared with an instructor-based reference diagram. The source code of DUET and supplementary materials are available in a GitHub repository\footnote{\url{https://github.com/sguertl/learnersourcing2025-duet}}.

\subsection{Architecture}
\begin{figure}
    \centering    \includegraphics[width=.75\linewidth]{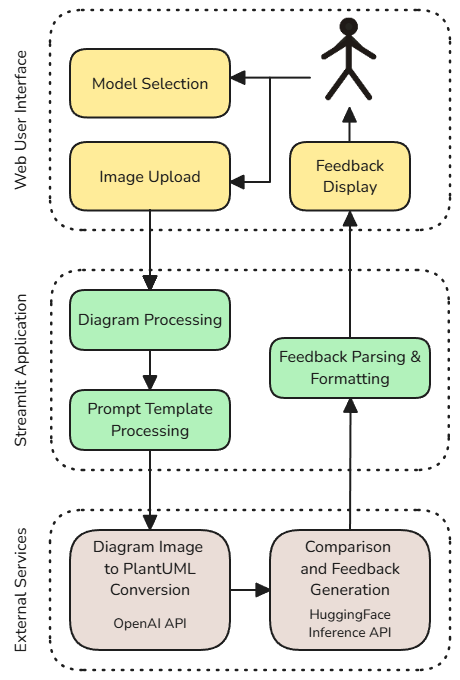}
    \caption{High-level system architecture of the automated UML and ER diagram feedback tool DUET. The web interface handles LLM selections and diagram uploads. GPT-4o converts diagram images into PlantUML code. A smaller LLM compares the representations and generates structured feedback for students and educators.}
    \label{fig:duet-system-architecture}
\end{figure}

Figure~\ref{fig:duet-system-architecture} depicts an overview of DUET's system architecture. The automated feedback tool is developed using Python and Streamlit\footnote{\url{https://streamlit.io/}}, which provides a web-based graphical user interface. The service is hosted on HuggingFace Spaces\footnote{\url{https://huggingface.co/spaces}}.

External LLMs are integrated via APIs and form the backbone of the tool's pipeline. OpenAI's \texttt{GPT-4o} initially converts the uploaded diagrams into PlantUML code. Users authenticate their access by entering a personal OpenAI API key. Subsequently, a smaller LLM, such as \texttt{Mistral-7B-Instruct-v0.3}, analyzes and compares the PlantUML codes against an instructor-provided reference solution. The tool currently accesses this LLM through HuggingFace's inference API, authenticated by a personal API key. However, the architecture also allows easy adaptation to locally hosted LLMs via frameworks such as Ollama\footnote{\url{https://ollama.com/}}.

The system uses predefined prompt templates for UML and ER diagrams. This approach allows for easy adaptation of prompts without modifications to the underlying application logic.

\subsection{Workflow and User Interaction}\label{sec:workflow}
\begin{figure*}
    \centering
    \includegraphics[width=\linewidth]{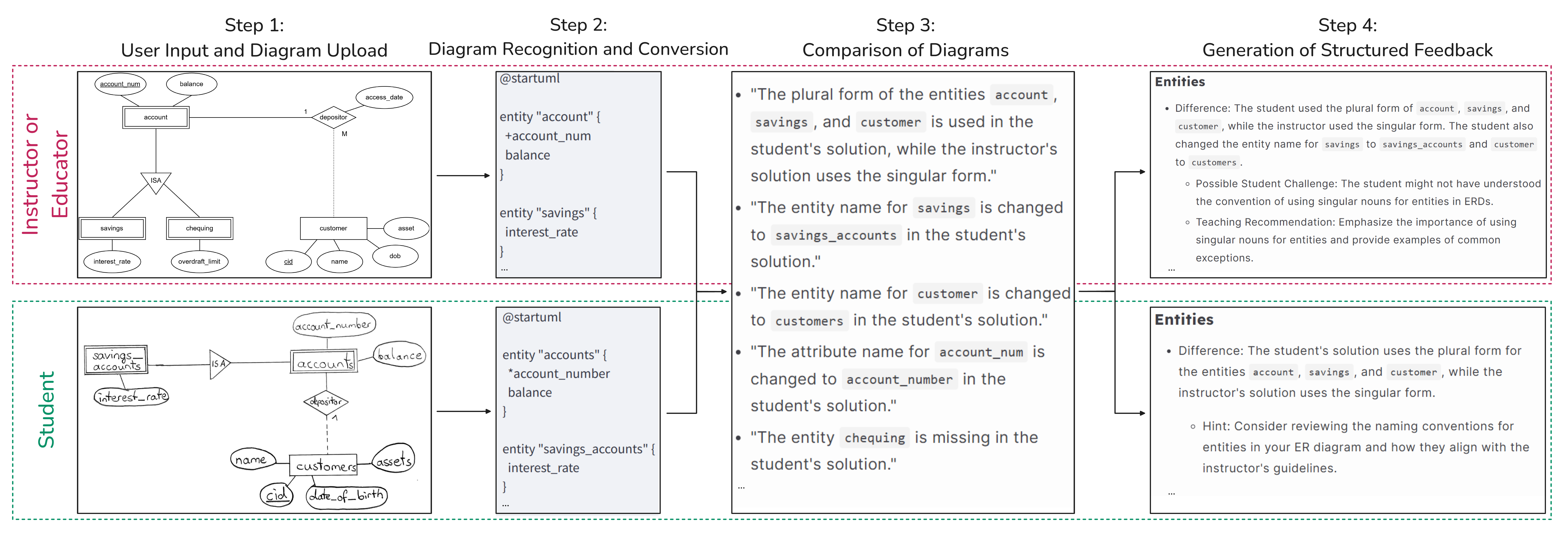}
    \caption{Illustration of DUET's workflow using ER diagrams, showing the process from diagram upload and conversion to PlantUML to comparison and structured feedback generation.}
    \label{fig:duet-example}
\end{figure*}

Building upon the system architecture, we illustrate the pipeline in Figure~\ref{fig:duet-example}, showing how users can interact with DUET and how diagrams are processed.\\ 

\textbf{Step 1: User Input and Diagram Upload.} Users select a language model for diagram recognition and another model for comparison and feedback generation. Once the model preferences are set, two diagrams can be uploaded: one serving as the instructor-provided reference solution and one representing the student's solution. The tool accepts both hand-drawn and digitally created UML and ER diagrams. In the current version, only a single student solution can be uploaded at a time, as bulk uploads are not yet supported.

Once both diagrams are submitted, users are guided to the next step.

\textbf{Step 2: Diagram Recognition and Conversion.} The tool processes both uploaded diagrams to extract their structural representations. OpenAI's \texttt{GPT-4o} converts the diagram images into PlantUML code. This step involves recognizing and interpreting visual elements such as classes, attributes, operations, and relationships. Since the recognition process depends on an LLM, the quality of the extracted PlantUML code may vary based on image quality, diagram clarity, handwriting legibility (for hand-drawn diagrams), and adherence to standard notations. The generated PlantUML code is the basis for the subsequent comparison and feedback generation.

\textbf{Step 3: Comparison of Diagrams.} After the conversion step, the tool compares the student's version against the instructor's reference solution. In this step, a smaller LLM, such as \texttt{Mistral-7B}, analyzes both representations to identify structural differences. The comparison focuses on objective elements such as missing or additional classes or entities, modified attributes, relationships, or changes in visibility. However, this step does not include any judgment or adjustments -- rather, it generates a list of differences that serves as the foundation for feedback generation in the next step.

\textbf{Step 4: Generation of Structured Feedback.} Based on the identified differences, the tool generates two types of structured feedback, again using a smaller LLM. The first type is directed at students and consists of reflective, neutral hints that encourage them to revisit specific aspects of their diagram without explicitly labeling elements as correct or incorrect. The second type targets educators and offers insights into potential student misunderstandings, misconceptions, and possible instructional implications. Both types of feedback are organized into predefined categories such as classes, attributes, and relationships. The generated feedback is displayed in the web interface as structured Markdown text, organized by predefined categories.

The current version of DUET is intended as a functional prototype for initial pilot testing in educational settings. It provides a foundation for iterative refinement and further adaptation based on feedback from both students and educators.

\section{Evaluation}

\subsection{Methods}
We conducted semi-structured interviews with six participants actively engaged in teaching courses frequently using UML and ER diagrams in the winter semester of 2024.
Two participants are researchers in computer science education and university lecturers. We refer to them as L1 and L2. L1 teaches a third-year \textit{Introduction to Databases} course for undergraduate computing majors and minors ($N = 300$) at a research-focused North American university, while L2 teaches a second-semester 
\textit{Introduction to Object-Oriented Programming} course for undergraduate computing majors ($N = 500$) at a research-focused European university. 
Furthermore, interviews were conducted with four senior graduate student TAs, with two TAs affiliated with each educator. These participants are referred to as TA1--TA4.

Each interview session took approximately 30 minutes and began with an introduction to the tool and a demonstration of its workflow, detailed in Section \ref{sec:workflow}. Subsequently, a discussion was held based on the questions in Table \ref{table:interview_questions}.

\begin{table}[h]
\caption{Questions asked at the evaluation interviews to six participants to identify strengths and weaknesses of DUET.}
\centering
\tabcolsep=0.21cm
\begin{tabular}{|c|p{7.5cm}|}
\hline
\textbf{Q\#} &  \textbf{Question} \\ \hline\hline
Q1 &  List and describe potential use cases for this tool?  \\\hline 
Q2 &  What features do you believe need to be refined?  \\\hline 
Q3 &  What features do you believe need to be added? \\\hline 
\multirow{3}{*}{Q4} &  In terms of using this tool in an introduction to databases or introduction to software engineering course, so think of an educational setting, what are 2-3 \textit{pros} of this tool? \\\hline
\multirow{3}{*}{Q5} &  In terms of using this tool in an introduction to databases or introduction to software engineering course, so think of an educational setting, what are 2-3 \textit{cons} of this tool? \\\hline 
Q6 &  Is there anything else you want to add or say?  \\ 
  \hline
\end{tabular}
    \label{table:interview_questions}
\end{table}

Two interviewers recorded notes in bullet-point format for each question and participants verified the notes for accuracy.
The collected interview notes were anonymized. 
We followed the thematic analysis approach, whereby segments of the notes were labeled to capture their essence, grouped into themes, and summarized. All data can be found in the project's GitHub repository. 

\subsection{Evaluation Results}
Four major categories emerged from the participants' responses in the summarized interview data.

\textbf{Benefits.} 
The six participants emphasized the tool's ability to provide instant and iterative feedback as a personalized, real-time learning aid, allowing students to correct errors and improve their understanding. 
L2 highlighted that ``\emph{students can use the system whenever they want and are not limited by the availability of TAs or lectures}''. Compared to traditional fixed-time lectures or labs, students can work on their assignments when it fits their schedule and receive immediate support.
Its availability and scalability offer easy adoption to large-scale courses without increasing the workload for tutors. TA1 and TA2 even saw the potential of reducing the workload for TAs and increasing fairness by enabling faster assessment using the tool's standardized feedback.

\textbf{Limitations.}
L1, TA1, TA3, and TA4 reported concerns about the accuracy and reliability of the tool. 
Specifically, the tool's non-deterministic output, resulting in variable and potentially inaccurate feedback across executions, can potentially lead to student confusion and avoidance of using the tool. 
Two participants (TA3, TA4) identified a possible way to cheat by generating solutions rather than elaborating on them. The possibility of tool misuse must be considered when implementing it in classroom settings.

\textbf{Use Cases in Learning.}
Instant feedback and an arbitrary number of interactions make the tool an effective learning asset. However, TA3 raised concerns about using it for assessment because of its non-deterministic output: ``\emph{Learners and educators can never be sure about the correctness, which makes it not suitable for assessment}''. 
To mitigate this concern, the tool should not be used to assess students' submissions automatically but rather as a learning tool to improve and learn interactively, with the information that results might differ for each run. 
L2, TA3, and TA4 emphasized that the tool should not replace TAs but offer another possibility to practice when no TA is available.

\textbf{Additional Feature Ideas.}
In addition to the benefits of the system for students, five participants (L1, L2, TA1, TA2, TA4) suggested that educators could gain additional insights into the learning process. L1 expressed: ``\emph{The tool could store all submission data to improve understanding of misconceptions through aggregated student submissions}''. Teaching methods could be adopted, and common errors could be explicitly discussed. 
Furthermore, TA3 and TA4 proposed a more interactive component, allowing students to ask questions after receiving feedback and to resolve misconceptions or misinterpretations. 
    
\section{Conclusion}
In this paper, we explore the use of LLMs to provide structured and automated feedback on UML and ER diagrams from the perspective of introductory software engineering and databases courses. The approach leverages LLM-based analysis to identify differences between student-created diagrams and instructors' reference solutions. Consequently, it enables reflective learning and targeted instructional interventions. To demonstrate this approach, we developed DUET, a prototype tool that applies LLMs to the analysis of student and instructor diagrams. The system integrates sophisticated models such as OpenAI's \texttt{GPT-4o} to convert diagram images into PlantUML code, and smaller models such as \texttt{Mistral-7B} to identify structural differences and generate feedback presented separately for students and educators.
Preliminary insights from semi-structured interviews with two educators and four TAs highlight initial use cases, perceived advantages, and potential limitations in applying this approach.

The use of LLMs to generate structured, formative feedback has the potential to support reflective learning and iterative improvement in introductory modeling courses in computing disciplines. As emphasized by the interviewees, constructive feedback could help students reconsider their design decisions without judging them as correct or incorrect, which could encourage deeper engagement with core concepts. In addition, immediate feedback may reduce students' reliance on instructor availability and enable them to receive and refine their work more independently. However, this approach and the tool are not intended to replace instructors or TAs, but rather serve as a supplementary component that supports and enhances existing teaching practices. For educators, this approach offers an opportunity to identify common misconceptions in student submissions, which are often challenging to detect systematically through manual review, and to inform targeted instructional interventions. In particular, in large-scale courses, automated feedback could help scale formative assessment while reducing instructor workload. For example, students might complete a supportive self-check with DUET before submitting their final diagram. This would allow students to reflect on possible structural issues and revise their work without interacting with instructors and thus depend on their availability. Alternatively, instructors could encourage students to use the tool during a preparatory activity before lab or tutorial sessions. This may help students arrive with more refined questions and a clearer understanding of their design choices. In addition, DUET may serve as a starting point for peer learning activities, where students collaboratively discuss the feedback received and subsequently compare alternative design strategies.

The results raise concerns about the accuracy and interpretability of LLM-generated feedback, particularly when dealing with complex or ambiguous diagrams. Since the feedback generation process remains largely a black box, it may be difficult for students and educators to fully understand or evaluate the reasoning behind the tool's hints and suggestions. The interviewees also noted the risk that students might rely too heavily on automated feedback and subsequent guidance, which could hinder the development of their modeling skills. Without opportunities to engage in problem-solving and self-correction, students may become passive recipients of feedback rather than actively developing their conceptual understanding and design reasoning. From a methodological perspective, the study involved a small number of participants and hypothetical scenarios, limiting the generalizability of the findings to real classroom contexts.

Future work will focus on improving DUET and deploying this approach in actual educational situations to evaluate its impact on student learning, instructional practices, and feedback effectiveness. Further development of the tool will explore improved system and user prompts, feedback presentation, and support for multiple student submissions and reusable reference solutions. Additionally, measures will be implemented to reduce the risk of misuse, such as using the tool for solution generation.

\section{Acknowledgments}
The authors thank the Natural Sciences and Engineering Research Council of Canada (NSERC), Discovery Grant \#RGPIN-2024-04348, for their financial support.

\bibliographystyle{ACM-Reference-Format}
\bibliography{references}

\end{document}